\documentclass[twocolumn,showpacs,preprintnumbers,amsmath,amssymb]{revtex4}
\usepackage{graphicx}
\usepackage{dcolumn}
\usepackage{bm}

\begin{document}

\title{Attempt at Perfecting Non-Relativistic Quantum Mechanics Based on Interaction}
\author{Tian-Hai Zeng}
 \email{zengtianhai@bit.edu.cn}

 \affiliation{School of Physics, Beijing Institute of Technology, Beijing 100081, People's Republic of China}%

\date{\today}
\begin{abstract}
  \textbf{Abstract} Many wave phenomena are related to interactions. Considering once neglected interactions in some cases, states of large objects and Newton's idea about measurement, we attempt to modify some concepts and principles of non-relativistic quantum mechanics. Our modifications may help one to understand some typical quantum phenomena with near classical and intuitive ideas without changing past correct results, may mediate some contradictions logically, and then may partially perfect non-relativistic quantum mechanics. We also give a review of the argument between Einstein and Bohr about non-relativistic quantum mechanics simply based on interaction.
\end{abstract}

\maketitle

\section{Introduction}

Scientists have made many great achievements in many fields by the application of non-relativistic quantum mechanics (NQM). These show the validity of the theory of NQM. Some things, however, such as the famous argument between Einstein \textit{et al.} \cite{A. Einstein} and Bohr \cite{Bohr} about the complete properties of NQM, and Feynman's assertion that no one can understand it \cite{R. P. Feynman}, motivate some people to try to understand and complete it. But a complete theory of NQM that Einstein believed and the wide recognition of some understanding of NQM do not yet appear until now.

No free particle exists. A particle always interacts with its environment, and then we have to consider that it is a subsystem of the composite system (CS) composed of it and its environment. When we talk about a free particle, it is an ideal concept of omitting any interaction (INT).

An interaction energy belongs to a composite system, which is composed of two or more subsystems, and does not belong to any subsystem. It is one of the four fundamental INT energies, \textit{i.e.}, gravitational, electromagnetic, strong and weak ones, or one composite within the four INTs. The external potential energy of a system is always the approximation of an INT energy.

When we know the expression of the INT energy in Hamiltonian between subsystems in a CS, we can solve the Schr\"{o}dinger equation in principle and obtain the knowledge of the evolution of the state and physical quantities of the CS.

An anti-intuitive quantum phenomenon may result from the neglect of the INT of the considered system with its environment and the neglect of the change of the state of the environment, due to the INT may be weak, the expression of the INT energy is unknown and the state of the environment is changing very slowly.

For examples, the INTs of a particle with a macro-double-slit or a splitter are unknown, the changes of some objects, which offer a potential barrier in a phenomenon of tunneling, were neglected.

So looking for unknown expressions of INT energies and giving out the kinetic operator form of the environment of a considered system are the key for us to understand quantum phenomena, and then to understand NQM, rather than looking for some hidden variable \cite{D. Bohm} that it was guessed might allow us to complete and understand NQM.

However, looking for unknown expressions of INT energies and the operator form of environment, and solving Schr\"{o}dinger equations are not the task of this article. We assume that the INTs in all CSs and the corresponding evolutions of states here are known.

By observations of some wave phenomena, for example, sound waves, water or liquid waves, or elastic waves in solids, we find a common point that any one of these waves has at least one INT between particles. But the wave equation, for example, a sound wave equation \cite{R. P. Feynman} ${\partial^2\xi}/\partial x^2=(1/\nu^2)\partial^2\xi/\partial t^2$ (here $\nu$ is the wave speed) may easily lead one to forget the existence of INT, for it does not appear clearly in the equation. The essence of superposition of any one of these waves is actually the superposition of INTs with the same particle, which induces interference phenomena. If there is no INT, these wave phenomena will not appear \cite{T. H. Zeng}.

When a large number of single particles go through a double-slit one by one, an interference phenomenon appears. Since there is no INT among single particles, single particles are considered to have an inherent probability wave property, \textit{i.e.}, every particle is in some superposed state after it goes through the double-slit. But there certainly exists some INT between a particle and the double-slit, in which case the superposed state of a particle and the interference phenomenon should be related with the INT and the state of the double-slit. We will explain this in Sec. 5 below entitled \textquotedblleft Understanding of typical quantum phenomena".

In Sec. 2, we offer a way of preparation of a superposed state starting from an eigenstate.  In Sec. 3, we give out modifications of NQM and some explanations. In Sec. 4, we suggest three laws correspond to Newton's three laws. In Sec. 6, we discuss some contradictions and mediations. In Sec. 7, we review the argument between Einstein and Bohr about physical reality and locality. The conclusion is in Sec. 8.

\section{Prepare a Superposed State Starting from an Eigenstate}

The starting point here is the consideration, how a particle state returns to the superposed state (or a wave packet state, expressed by a wave function or state vector), after its initial superposed state has collapsed to an eigenstate (eigenfunction).

A momentum eigenstate of a free particle never automatically evolves into a superposed state of different momentum eigenstates and its momentum eigenvalue remains unchanged in its evolution according to the Schr\"{o}dinger equation. A general way of preparing a particle in a superposed state with different momentum eigenstates from its initial momentum eigenstate, should let it interact with an object (apparatus or environment), and then it can exchange momentum with the object.

We consider preparing a particle in a superposed state with different momentum eigenstates (they are also kinetic energy eigenstates) from its initial one of its momentum eigenstates, expressed by state vector $|\vec{p}_1\rangle$. The apparatus is initially in the momentum eigenstate $|\vec{P}_1\rangle$.

After switching the INT between the two parts, the two parts exchange momenta and the CS evolves usually into an entangled state, which was first named by Schr\"{o}dinger \cite{J. A. Wheeler}. The definition of entangled state is that it cannot be expressed in the factoring form or product state of the states of parts in a composite system. A product state, for example, $|\phi\rangle|\Phi\rangle$, is the product of a state $|\phi\rangle$ of the particle and a state $|\Phi\rangle$ of the apparatus. According to the Schr\"{o}dinger equation, we may obtain in principle one entangled state for simplicity in the form:
\begin{equation}\label{l}
|\Psi\rangle=|\vec{p}_1\rangle|\vec{P}_1\rangle+|\vec{p}_2\rangle|\vec{P}_2\rangle,
\end{equation}
where $|\vec{p}_2\rangle$ and $|\vec{P}_2\rangle$ are other momentum eigenstates for the particle and the apparatus, respectively, and the normalizing factors are often neglected in this article. The familiar example is hydrogen atom, the state of the composite system of the electron and the proton is an entangled one whatever state the atom stays.

If the apparatus is much larger than the particle, the states can be considered as the approximately same $|\vec{P}_1\rangle\approx|\vec{P}_2\rangle$ (the magnitudes and directions of the two momentum eigenvalues) in the considered short evolution time, and then we can obtain an approximate product state:
\begin{equation}\label{2}
|\vec{p}_1\rangle|\vec{P}_1\rangle+|\vec{p}_2\rangle|\vec{P}_2\rangle\approx(|\vec{p}_1\rangle+|\vec{p}_2\rangle)|\vec{P}_1\rangle,
\end{equation}
The two parts are approximately separate, and so we prepare the particle in an approximately superposed state $|\vec{p}_1\rangle+|\vec{p}_2\rangle$ of different momentum eigenstates by an apparatus \cite{T. H. Zeng, M. Fayngold}.

We fail to find any other way to obtain the superposed state of momentum of a particle. According to NQM, if we measure the momentum of the apparatus in the base $\{|\vec{P}_1\rangle+|\vec{P}_2\rangle$, $|\vec{P}_1\rangle-|\vec{P}_2\rangle\}$, we seem to get a superposed state $|\vec{p}_1\rangle+|\vec{p}_2\rangle$ or $|\vec{p}_1\rangle-|\vec{p}_2\rangle$ from Eq. (1), but no such base (two superposed states) of a macro-apparatus exists. If the apparatus is a micro-object, we seem to be able to measure the momentum of the apparatus in the base, or to use an eraser for a delayed-choice \cite{J. S. Tang}, and then to get a superposed state of the particle, but this is a pre-hypothesis that the two superposed states of an apparatus have existed, which contradicts the premise of preparing a particle or a micro-object in a superposed state.

Tracing out the states of the apparatus in Eq. (1), or measuring the momentum of the apparatus but not reading the result, we only obtain a mixed state $|\vec{p}_1\rangle\langle\vec{p}_1|+|\vec{p}_2\rangle\langle\vec{p}_2|$   (without coherence) of the particle. According to NQM, the entanglement still holds even after the INT ceases, so the approximation above may be the final step to obtain a superposed state of a particle with different momentum eigenstates, and we cannot prepare a free particle (without interacting and entangling with another object) in the pure superposed state $|\vec{p}_1\rangle+|\vec{p}_2\rangle$.

Similar with $|\vec{P}_1\rangle\approx|\vec{P}_2\rangle$, Bertet \textit{et al.} \cite{P. Bertet} considered that the coherent state $|\beta_g\rangle\approx|\beta_e\rangle$ in the classical limit. But they have not written their equation (3) as $(|g\rangle|\beta_g\rangle+|e\rangle|\beta_e\rangle)/\sqrt{2}\approx(|g\rangle+|e\rangle)|\beta_g\rangle/\sqrt{2}$, and have not obtained the approximately superposed state of an atom with two levels $|g\rangle$ and $|e\rangle$.

Some physical quantities are exchanged between entangled parts by their INTs. If the INT ceases, then no momentum (or other physical quantities) is exchanged between the particle and the apparatus in Eq. (1), so the entangled state is disentangled and no such superposed state $|\vec{p}_1\rangle+|\vec{p}_2\rangle$ of a free particle can be obtained. These examples explain that the entangled states and the approximately superposed states are determined by the Schr\"{o}dinger equation of CS including at least one INT. Can someone imagine any way to circumvent INT to obtain superposed states of a free particle?

It is reasonable to consider the INT between a particle and a macro-apparatus or environment similar to the case of decoherence \cite{W. H. Zurek}, in which the environment entangles with a quantum system, and then destroys the coherence in its superposed states due to unavoidable INT between them in a not short evolution time.

We suggest the mass ratio of proton/electron, 1836, or other constant in some specific case, being as a criterion of neglecting the change of state of one object in a CS, if the mass, or energy, or magnitude of momentum, \textit{etc.}, of the object is 1836 times a corresponding quantity of a particle, like the proton in a hydrogen atom in NQM. We call such an object a  \textit{large object} (LO). In a short evolution time, the change of an LO's state approximately does not affect a quantum system when they interact each other, although the INT always affects the quantum system.

\section{Modifications and Explanations}

On the concept of wave-particle duality in NQM, the wave and particle properties of a particle are equally intrinsic, and they are all independent of INT. Our modification is: Wave property is based on INT. An INT with an LO makes a particle behave with the property of superposition of probability waves or states, and then behave with the property of interference. Two or more particles, as a CS, have the property of superposition of probability waves or states due to interactions. If there is no INT, a free particle will not show the probability wave phenomena. A de Broglie wave vector may be considered as only a parameter corresponding to a definite momentum and energy of a free particle.

The principle of superposition of states of NQM is that a superposed state of two or more different states of a particle (or system) is still its state, and that it is also independent of INT. Our modification is: Superposed states exist only in CSs with INTs between subsystems, they are entangled states, and INTs and the conservation laws (below) are restrictive conditions. If the state of a particle or a system can be expanded into a superposed state of different eigenstates, such as energy, momentum and angular momentum eigenstates, etc., it must be an approximate one due to the fact that an INT with an LO lets them evolve into an entangled state, and the state of the LO is approximately unchanged in considered short evolution time. This has been explained in the above examples.

The Schr\"{o}dinger equation in NQM is a dynamic equation describing a CS, or a particle in a field, even a free particle or an isolated system in a superposed state with different energy eigenstates. Our modification is: The Schr\"{o}dinger equation is a dynamic equation that strictly describes a CS with INTs between subsystems or particles, and the wave function must include all subsystems, even including an LO or environment. For example, an equation of a CS of two subsystems is:

 \begin{equation}\label{3}
i\hbar\frac{\partial\Psi}{\partial t}=(\frac{\hat{\overrightarrow{p_{1}}} ^{2}}{2m_{1}}+V(r_{12})+\frac{\hat{\overrightarrow{p_{2}}} ^{2}}{2m_{2}})\Psi,
\end{equation}
where ${\hat{\overrightarrow{p_{1}}} ^{2}}/{2m_{1}}$ and ${\hat{\overrightarrow{p_{2}}} ^{2}}/{2m_{2}}$ are the kinetic energy operators of two micro-subsystems, or the latter ${\hat{\overrightarrow{p_{2}}} ^{2}}/{2m_{2}}$ is formally that of an LO, and $V(r_{12})$ is the INT (potential) energy belonging to the CS.

We think that any external field for a particle (or a micro-system) is the approximation the interaction energy between the particle and its environment.

The equation of a particle or a micro-system in an external field must be the approximation of Eq. (3), in which ${\hat{\overrightarrow{p_{2}}} ^{2}}/{2m_{2}}$ and the state of an LO in $\Psi$ have been neglected, for the state is approximately unchanged in the considered evolution time. The INT has been approximately considered as an external field and the INT energy is considered as belonging to the micro-system.

The Heisenberg equation and the Feynman path integral often describe the behaviour of a particle or a micro-system, therefore they are approximate expressions.

The Schr\"{o}dinger equation is also formally a kinematic equation of a free particle or an isolated quantum system, with definite energy, momentum and angular momentum, for there are no exchanges of these quantities with other object.

Conservation laws in NQM are only of the sense of statistical average. Our modification is: In a CS, the three physical quantities, energy, momentum and angular momentum, are exchanged among subsystems by INTs and strictly maintain respective conservations at any time (or in a single process) and naturally in the sense of the statistical average.

Some experimental results negated the point of Bohr \textit{et al}. \cite{N. Bohr} that the conservation laws in the micro-world hold only in the sense of the statistical average. For examples, in 1925, W. Bothe and H. Geiger \cite{W. Bothe,  Bothe} by single Compton collisions, and in 1932, J. Chadwick's discovery of the neutron \cite{J. Chadwick}, proved experimentally that the conservation laws of energy and momentum hold in single processes.

For the state $|\vec{p}_1\rangle|\vec{P}_1\rangle+|\vec{p}_2\rangle|\vec{P}_2\rangle$ in Eq. (1), the INT exchanges the momentum difference $\vec{p}_1-\vec{p}_2=\vec{P}_2-\vec{P}_1$ between the two parts, and then $\vec{p}_1+\vec{P}_1=\vec{p}_2+\vec{P}_2$. The total momentum is conserved at any time whatever the state collapses into $|\vec{p}_1\rangle|\vec{P}_1\rangle$ or $|\vec{p}_2\rangle|\vec{P}_2\rangle$. These terms are degenerate for momentum. If $\vec{p}_1+\vec{P}_1\neq\vec{p}_2+\vec{P}_2$, there must be at least one LO or environment interacting with the two parts and the three parts evolve into an entangled state, and the momenta $\vec{M}_1$ and $\vec{M}_2$ of the LO should meet $\vec{p}_1+\vec{P}_1+\vec{M}_1=\vec{p}_2+\vec{P}_2+\vec{M}_2$ with $|\vec{M}_1\rangle\approx|\vec{M}_2\rangle$. When the state is expanded by angular momentum eigenstates, it must be of the form $|\vec{l}_1\rangle|\vec{L}_1\rangle+|\vec{l}_2\rangle|\vec{L}_2\rangle+\cdot\cdot\cdot+|\vec{l}_N\rangle|\vec{L}_N\rangle$ with $\vec{l}_1+\vec{L}_1=\vec{l}_2+\vec{L}_2=\cdot\cdot\cdot=\vec{l}_N+\vec{L}_N$, \textit{i.e.}, the conservation law of angular momentum holds at any time.

The conservation law of energy is a bit more complex than the conservation laws of momentum and angular momentum, due to interaction (INT) energy belonging to the whole CS, \textit{i.e.}, not belonging to one subsystem. Given a particle (or a micro-system) initially in one of its kinetic energy eigenstates $|k_1\rangle$ with kinetic energy $k_1$, and an apparatus also initially in one of its kinetic energy eigenstates $|K_1\rangle$ with kinetic energy $K_1$, after switching the INT between them, the CS evolves into a state $|\varepsilon_1\rangle+|\varepsilon_2\rangle$ for some INT for simplicity, and the total energy is
\begin{equation}\label{4}
\varepsilon_1=k_1+K_1+V_1=k_2+K_2+V_2=\varepsilon_2,
\end{equation}
in which the energy conservation holds, where $V_1$ and $V_2$ are INT energies, and $k_2$ and $K_2$ are kinetic energies of the particle and the apparatus, respectively. If the apparatus is an LO, then $K_1\approx K_2$ and the INT energy can be considered approximately to belong to the particle. The particle energy is $en_\emph{i}=k_\emph{i}+V_\emph{i}$, $\emph{i}=1,2$, and then the state is an entangled state $|en_1\rangle|K_1\rangle+|en_2\rangle|K_2\rangle\approx(|en_1\rangle+|en_2\rangle)|K_1\rangle$, where $|en_1\rangle$ and $|en_2\rangle$ are its energy eigenstates. We believe that the weaker the interaction between a CS with its environment is, the more strictly the conservation laws for the CS hold experimentally.

From the conservation laws, we can deduce that when INTs in a CS cease, \emph{i.e.}, there is no exchange of the three physical quantities between the subsystems, the subsystems must be in their respective eigenstates with definite energy, momentum and angular momentum, and then the CS is in a product state, \emph{i.e.}, no longer in an entangled state.

Here quantum measurement is limited to the measurement of micro-objects, not that of macro-objects reaching accuracy of the order of the standard quantum limit  \cite{V. B. Braginsky}. When an apparatus interacts with a measured quantum system and their respective states are all changed, the result of the measurement can be obtained from the change of the apparatus state. If the system state is unchanged, the apparatus state should be unchanged yet, and this case yet corresponds to no measurement, so that no result or information about the system can be read out.

This real measurement of micro-systems is different from the measurement of macro-systems, in which the result can be obtained without disturbing the measured macro-system in many cases where only \textit{micro-action} exists; for example, measuring the length of a macro-object may neglect some photon-action on the two ends of the object. This point is close to Newton's idea \cite{Newton} that \textquotedblleft Relative quantities are not the actual quantities whose names they bear but are those sensible measures of them (whether true or erroneous) that are commonly used instead of the quantities being measured". This is suitable for directly measured (below) physical quantities of quantum systems.

However, the depiction of quantum measurement in most NQM books is simple. It is same as that of Dirac \cite{P. A. M. Dirac}. The measurement result is an eigenvalue with some probability if the initial state of a measured system is not an eigenstate of the measured quantity, and the state left is an eigenstate corresponding to the eigenvalue.

Our modification is complex. We may vaguely divide the measurement process into two steps: operating process and reading out result. We divide the quantum measurement ways to two types: non-direct measurement and direct measurement.

Non-direct measurement is to change \emph{A} ( representing one or more physical quantities of the measured system), which is the operating process, and then to read out the result, the eigenvalue of some quantity \emph{B}, from the change of measuring apparatus. This is different from indirect measurement in \cite{V. B. Braginsky}. If reading out result does not destroy the final eigenstate corresponding to the eigenvalue, the non-direct measurement is same as the prediction of Dirac's book \cite{P. A. M. Dirac} and most NQM books. However, there exists some case in which the eigenstate corresponding to the read out eigenvalue is destroyed.

For example, the eigenvalue of the spin state of an electron (or the polarization of a photon) can be non-directly measured by changing the direction of the electron's motion in the inhomogeneous magnetic field in the Stern-Gerlach experiment. If the electron clicks a detector, the spin eigenstate corresponding to the read out spin eigenvalue is destroyed. If there is no detector and the electron separates from (or ceases the interaction with) the Stern-Gerlach apparatus or the magnetic field, the electron in its appearance place will stay in the spin eigenstate with a definite spin eigenvalue.

Direct measurement is to change the measured physical quantity \emph{C} of the measured system, and then to read out the difference of the eigenvalue of its final eigenstate and one of probable eigenvalues of its initial state from the change of measuring apparatus. This is different from the same term, direct measurement, in \cite{V. B. Braginsky}.

In the experiment of absorption spectrum, for example, an atom at ground level absorbs a photon and its level jumps into an excited one, so the result of the measurement of its energy is then the difference of the two levels, equal to the energy of the photon, and the atom is in the excited level after measurement, rather than remains its initial energy eigenstate or level unchanged.

The difference of two eigenvalues of a physical quantity of the measured system is not an error. The real error of a measurement result comes from the process of amplifying the signal in apparatus.

We may obtain the expectation value of the above direct measurement. For example, if a particle is initially in an energy eigenstate $|E_0\rangle$ (assuming $E_0=0$), the final state is some mixed state $\rho_{f}=\Sigma P_{n} |E_{n}\rangle \langle E_{n}|$ (assuming all $E_{n}>E_0$) after measurement, and then the expectation value of energy is
\begin{equation}\label{5}
\langle E \rangle=\Sigma P_{n} E_{n}
\end{equation}

Two uncertainty relations in NQM are $\triangle p \triangle x\geq\hbar/2$ and $\triangle E \triangle t\geq\hbar/2$. The explanation of the former is that the greater the accuracy with which the coordinate \emph{x} of a particle is known (\textit{i.e.}, the less  $\triangle x$ is), the greater the momentum uncertainty  $\triangle p$ of the particle \cite{L. D. Landau} is. The explanation of the latter is that the smaller the time interval $\triangle t$ (of two measurements) is, the greater the energy change that is observed.

Our modification is: The least change of angular momentum is $h$ (Planck constant), which is the least exchanging action between one particle and another one or an apparatus; and a change of the energy or the momentum of a measured particle is equal to the whole or part (Compton collisions) of the energy or momentum of one photon.

For example, if a photon interacts with an atom and is absorbed by the atom, the increased energy of the atom has been measured,
\begin{equation}\label{6}
\triangle E=h/T
\end{equation}
where \emph{T} is the inverse of the photon frequency and may be considered as a characteristic time of the photon being absorbed by the atom. Completing a measurement needs a time interval, $\triangle t\geq T$, then we obtain
\begin{equation}\label{7}
\triangle E \triangle t\geq h
\end{equation}
where \emph{h} is different from $\hbar/2$ in the Heisenberg uncertainty relation  and the $\triangle t$ is not a time interval of two measurements. Similarly, in \emph{x} direction the change of the momentum of the atom is also equal to that of the photon, $\triangle p=h/cT$, where \emph{c} is the speed of light. Completing a measurement also needs a space extent, $\triangle x\geq cT$, then $\triangle p \triangle x\geq h$.

The principle of the indistinguishability of identical particles in NQM is that the wave function of the system is either symmetrical or antisymmetrical, and the system is in a pure entangled state.

Our modification is: The system can be in a pure entangled state only if there exists INT between any pair of identical particles; when all INTs cease, any particle is free and is in an eigenstate having definite energy, momentum and angular momentum, and the pure entangled state of the system is disentangled and is only described as a mixed state. The indistinguishability fits for the mixed state as well as the pure entangled state.

\section{Three laws correspond to Newton's three laws}

We believe that the following three laws, in the micro-world discussed in non-relativistic quantum mechanics, correspond to Newton's three laws in classical mechanics.

The Schr\"{o}dinger equation, that strictly describes a CS with INTs between subsystems or particles, corresponds to Newton's second law. The Heisenberg equation and the Feynman path integral often describe a particle or a micro-system in an external field, they must be approximate as similar as the Schr\"{o}dinger equation of one particle.

When INTs cease, a free particle has a definite energy, momentum and angular momentum, and it is still described formally by the Schr\"{o}dinger equation, the Heisenberg equation and the Feynman path integral; this corresponds to Newton's first law, in which a macro-particle is still described formally by Newton's second law under the sum of all forces being zero.

The above modified principle of the superposition of states, that superposed states exist only in CSs with INTs between subsystems and are entangled states with INTs and the conservation laws being restrictive conditions, corresponds to Newton's third law.

\section{Understanding of Typical Quantum Phenomena}

The most typical quantum phenomenon is double-slit interference of material particles. The general depiction of the interference of double-slit is such that, we let particles go one by one through double-slit and obtain an interference pattern on a distant screen; there is no INT between particles; some particles cannot go through the double-slit; if we detect particles just behind \cite{R. P. Feynman} or just before \cite{M. O. Scully} the slits and know their which-way information, the interference pattern will disappear; if we do not detect particles just behind or just before the slits for getting which-way information, or detect particles just before the slits and erase the which-way information \cite{M. O. Scully}, the interference pattern will appear on a distant screen; only a whole particle can be detected at a time in one of the two detectors just behind the two slits, and no parts of the particle can be detected.

The general explanation is that, a particle in a form of wave goes through the two slits, and then the state of the wave or the particle is the superposition of the two path states through the two slits, $|\psi_1\rangle+|\psi_2\rangle$, which has coherence; all particles through the two slits stay in the same state and reach some point on the distant screen with a definite probability, and then they form an interference pattern.

The general explanation of the interference of a macro-double-slit is not related with the INT between a particle and the matter of the macro-double-slit. If the INT does not exist, we will not know how a particle from a plan wave state evolves into the superposition of two path states according to Schr\"{o}dinger equation.

In recent papers, however, the authors have to consider the INT between a particle and the matter of a micro-double-slit in their calculations, and obtained the results agreeing with their experimental ones. These papers can be divided into two cases.

The first case \cite{D. Akoury, R.K. Kushawaha} is that the matter of a micro-double-slit (two protons in $H_{2}$ or two carbon nuclei in hydrocarbons) is larger than a projectile particle (electron). Their calculations include the INT between a projectile particle and the double-slit in Schr\"{o}dinger equation or Hamiltonian, but neglect the kinetic energy operator of the double-slit and then do not obtain the entangled state of the CS. The calculations directly obtain the coherent superposed state of a projectile particle, which is not an approximately superposed state from the entangled state with the state of the double-slit approximately unchanged.

The second case \cite{L. Ph. H. Schmidt} is that the micro-double-slit ($H_{2}^{+}$ ) is even smaller than a projectile particle $He$, their calculations also include the INT between $H_{2}^{+}$ and $He$.

It is reasonable that we consider the INT between a particle and a macro-double-slit, and the INT drives the CS to evolve into an entangled state according to Schr\"{o}dinger equation in principle, though the form of the INT is unknown. Therefore, we may offer a new experiment way and a new explanation, without saying a particle through a double-slit in a wave form.

If the particle source aims at one slit, i.e., to turn the source a small angle or to move the source a small distance perpendicular to the plain of the source point and the middle line of the two-slit, a particle will have high probability to pass the slit, the INT between the particle and the matter of double-slit drives the CS to evolve into an entangled state according to the Schr\"{o}dinger equation in principle. Assume that the time evolution is
\begin{multline}\label{8}
  |\psi_1(0)\rangle|S_1(0)\rangle\rightarrow\\
  \alpha_1(t)|\psi_1(t)\rangle|S_1(t)\rangle+\beta_1(t)|\psi_2(t)\rangle|S_2(t)\rangle
\end{multline}
If the particle source aims at the other slit, the time evolution is
\begin{multline}\label{9}
  |\psi_2(0)\rangle|S_2(0)\rangle\rightarrow\\
 \alpha_2(t)|\psi_1(t)\rangle|S_1(t)\rangle +\beta_2(t)|\psi_2(t)\rangle|S_2(t)\rangle
\end{multline}
where $|\psi_1(0)\rangle$ ($|\psi_2(0)\rangle$) is the initial path state of a particle just passing through the slit 1 (slit 2) and $|S_1(0)\rangle$ ($|S_2(0)\rangle$) is the corresponding state of the matter of the double-slit, and the normalizing factors $\alpha_1$, $\beta_1$, $\alpha_2$, $\beta_2$ express the difference of the two entangled states. In symmetrical case, $|\alpha_1|=|\beta_2|>|\beta_1|=|\alpha_2|$.

The two entangled states can be approximately expressed in one form $(|\psi_1\rangle|S_1\rangle+|\psi_2\rangle|S_2\rangle)/\sqrt{2}$ after a suitable evolution time (or passing a suitable distance), \textit{i.e.}, the time or the distance of a particle from the double-slit to a screen, which is placed such that interference patterns can be observed. Since the state $|S_1\rangle\approx|S_2\rangle$ within the evolution time, we can obtain an approximately superposed state $(|\psi_1\rangle+|\psi_2\rangle)/\sqrt{2}$ of the particle due to the INT.

If the time or the distance is short, the interference pattern will be the addition of the two interference patterns of the two states $\alpha_1|\psi_1\rangle+\beta_1|\psi_2\rangle$ and $\alpha_2|\psi_1\rangle+\beta_2|\psi_2\rangle$. Then the interference pattern changes from with two peaks to general pattern when the distance from the screen to the double-slit increases \cite{S. Kocsis}.

The phenomena of particles tunneling through a barrier region or being reflected by a potential well with some probability are considered as pure quantum effects and no one can understand them \cite{D. J. Griffiths}.

According to our modification, when a particle enters the barrier region, the particle must interact with some LO and the energy of the potential barrier belongs to the CS, and then the CS evolves into an entangled state according to the Schr\"{o}dinger equation.

We use $|a_2\rangle$ and $|a_3\rangle$ expressing the two energy eigenstates of the particle in the barrier region for simplicity, where the energy in $|a_3\rangle$ is greater than that of the barrier due to INT (the particle may get some energy from the LO), and use $|a_1\rangle$ and $|a_4\rangle$ expressing two energy eigenstates before and behind the region, respectively. The corresponding states of the LO are approximately unchanged in a short time and so the entangled state may be approximately expressed as $(c_1|a_1\rangle+c_2|a_2\rangle+c_3|a_3\rangle+c_4|a_4\rangle)|LO\rangle$. Therefore, the particle may be found in the three regions with probabilities $|c_1|^{2}$, $|c_2|^{2}+|c_3|^{2}$, and $|c_4|^{2}$.

The consideration of INT potential energy belonging to a particle sometimes may bring an absurd understanding. In the case of describing a particle by the Schr\"{o}dinger equation, the potential energy term in Hamiltonian is often considered as belonging to the particle, so the energy of the particle includes kinetic and potential barrier energy. If a particle enters the barrier region, it seems that its energy can become greater than its initial kinetic energy due to the potential barrier energy belonging to it, and then it can naturally jump the barrier with a large probability.

One coherent state \cite{ R. J. Glauber} of optical field $|\alpha\rangle$ ($\alpha$ is an arbitrary complex number) may be expanded in terms of photon number states $|n\rangle$ as $|\alpha\rangle=\Sigma f(\alpha,n) |n\rangle$ ($n=0,1,2,\cdots,\infty$), where $f(\alpha,n)$ are coefficients. The interaction between the light source apparatus and the optical field always exists and drives the CS to evolve into an entangled state. The state of the apparatus can be considered as approximately unchanged in a short time, and then the entangled state can be an approximate product state and the coherent state can be explained as an approximately superposed state of number states.

In Ref.\cite{C. Monroe} C. Monroe \textit{et al}. prepared the \emph{Schr\"{o}dinger's cat}\cite{J. A. Wheeler} superposition state of an atom, $|x_{1}\rangle|\uparrow\rangle+|x_{2}\rangle|\downarrow\rangle$, an entangled state of its external space states and its internal energy states. It is actually the entangled state, $(|x_{1}\rangle|\uparrow\rangle|LO_1\rangle+|x_{2}\rangle|\downarrow\rangle|LO_2\rangle\approx(|x_{1}\rangle|\uparrow\rangle+|x_{2}\rangle|\downarrow\rangle)|LO_1\rangle$, of the CS of the atom and an LO, which is Raman beam and its state is approximately unchanged.

\section{Contradictions and Mediations}

The following two points of measurement contradict each other. Landau \textit{et al}. \cite{L. D. Landau} thought that: \textquotedblleft After the measurement, however, the electron is in a state different from its initial one, and in this state the quantity \emph{f} does not in general take any definite value. Hence, on carrying out a second measurement on the electron immediately after the first, we should obtain for \emph{f} a value which did not agree with that obtained from the first measurement." This contradicts the point made by Dirac \cite{P. A. M. Dirac} that: \textquotedblleft From physical continuity, if we make a second measurement of the same dynamical variable $\xi$ immediately after the first, the result of the second measurement must be the same as that of the first. Hence after the first measurement has been made, the system is in an eigenstate of the dynamical variable, the eigenvalue it belongs to being equal to the result of the first measurement."

According to our modification, If reading out result does not destroy the final eigenstate corresponding to the eigenvalue, the non-direct measurement obtains that the result of the second measurement must be the same as that of the first, while the value of at least one other quantity must be changed. For example, in the non-direct measuring the spin of an electron in Stern-Gerlach experiment, the motion direction of the electron is changing during the second experiment, but the spin is same as the first. This is close to Dirac's point.

On the other hand, the directly measured quantities of a particle, such as energy, momentum and angular momentum, \textit{etc}., only measure the difference between a final eigenvalue after the measurement and one of the eigenvalues of a superposed state before the measurement. On carrying out a second measurement on the same quantity immediately after the first, the two results may be different and the two states after two measurements may also be different. This is close to the point of Landau \textit{et al}., but directly measured the difference between two eigenvalues is different from Landau's point that the measured quantity is an eigenvalue. So we partly mediate the contradiction between them.

In NQM, a free atom can be in a superposed state $|g\rangle+|e\rangle$ of its ground state $|g\rangle$ and an excited state $|e\rangle$. A free electron (or one photon) can be in a superposed spin state $|s\rangle=|0\rangle+|1\rangle$ of its spin up state $|0\rangle$ (vertical polarized or path 0) and spin down state $|1\rangle$ (horizontal polarized or path 1). If the above states are expressed in the form $|live cat\rangle+|dead cat\rangle$, the cat should be even stranger than Schr\"{o}dinger's cat $|live cat\rangle|no toxicant\rangle+|dead cat\rangle|toxicant\rangle$. When we discover that Schr\"{o}dinger's cat is dead, the reason must be the toxicant killing it, but we cannot find any matter that might have killed the cat initially in the state $|live cat\rangle+|dead cat\rangle$.

According to our modification, when we discover the cat is dead, the reason must be that an INT with an LO killed it. There only exists an atom being in an approximately superposed state $|g\rangle+|e\rangle$ due to INT with an LO or apparatus. When an electron in some state $|s\rangle$ interacts with the non-uniform magnetic field $|SG\rangle$ of a Stern-Gerlach apparatus, the CS evolves into an entangled state $|s\rangle|SG\rangle\rightarrow|0\rangle|SG_1\rangle+|1\rangle|SG_2\rangle\approx(|0\rangle+|1\rangle)|SG_1\rangle$, where  $|SG_1\rangle$ and $|SG_2\rangle$ are states of the magnetic field after switching their interaction and are approximately the same, therefore the electron is in the approximately superposed state $|0\rangle+|1\rangle)$. A free electron in some state $|s\rangle$ must interact with an LO or apparatus, then the state $|s\rangle$ may be dissolved into the superposed state $|0\rangle+|1\rangle)$ in physics, though it is easily dissolved into a superposed state mathematically. A photon polarized state or a photon path state is similar. When the photon interacts with a beam splitter, its state may be dissolved into a superposed state of different polarized states or path states.

In NQM, measuring the energy of a free atom in state $|g\rangle+|e\rangle$ brings a collapsed state $|g\rangle$ or $|e\rangle$. Assume that some energy change introduced by an apparatus is much smaller than the difference $\Delta E$ of the two atomic levels; then the energy conservation does not strictly hold in the CS of the atom and the apparatus. A lifetime of the free atom state $|g\rangle+|e\rangle$ is $\Delta t$ due to $\Delta E\Delta t\sim\hbar/2$, and the state spontaneously collapses into state $|g\rangle$ or $|e\rangle$, so the atomic energy is also non-conservation. The non-conservation is reluctantly explained as only existing in the short time $\Delta t$, whereas if we explain this by selecting the relation $\Delta E\Delta t>\hbar/2$, then big energy non-conservation may exist for a long time, which is not acceptable. If the short lifetime $\Delta t$ is acceptable, it contradicts the principle that the evolution time of a free atom state   has no limit according to Schr\"{o}dinger's equation.

We explain that, a free atom is not in state $|g\rangle+|e\rangle$. Before and after a measurement of the atomic energy (initially being in the approximately superposed state $|g\rangle+|e\rangle$), the energy conservation strictly holds considering the interchanges of energies among the atom, apparatus and an LO. When we discuss the evolution time of an atomic state $|g\rangle+|e\rangle$, it is actually that of the CS of the atom and an LO. This evolution time has no limit according to the Schr\"{o}dinger equation or even according to the relation $\Delta E\Delta t\sim\hbar/2$ due to $\Delta E=0$, \textit{i.e}., the energy conservation still holds strictly. If an INT quantum (or photon) escapes, or (and) the state of an LO is evidently changed, the evolution of $|g\rangle+|e\rangle$ ends, and the system of the atom and a new LO start a new evolution.

It will produce another contradiction when we combine Dirac's point \cite{P. A. M. Dirac} and the uncertainty relation given by Landau \emph{et al.} \cite{L. D. Landau}, that $\Delta E$ is the difference of the two measured energy eigenvalues and $\Delta t$ is the interval of the two measurements. If the energy of an atom is measured twice in a finite interval $\Delta t$ and the same energy is obtained according to Dirac's point, then the difference  $\Delta E=0\rightarrow\Delta E\Delta t=0$, and this contradicts the Heisenberg uncertainty relation.

According to our modified uncertainty relation $\Delta E\Delta t\geq h$, $\Delta E$ is the difference of two different energy eigenvalues and $\Delta t$ is the time in one measurement (not the time interval of two measurements), and this then avoids the relation $\Delta E\Delta t=0$.

In NQM, after two identical particles separate distantly and the INT between them ceases, the entanglement of the initial state $|e_1\rangle|e_2\rangle+|e_2\rangle|e_1\rangle$ of the energy eigenstate still persists. This contradicts the idea that the energies of two particles should be exchanged with each other by INT. We can understand the state $|g\rangle|1\rangle+|e\rangle|0\rangle$, when $n=0$  in Eq. (3), but we cannot understand the state $|g\rangle|0\rangle+|e\rangle|1\rangle$ in NQM, due to not knowing how to exchange energy between the atom and the field.

We explain that, if there exists an LO interacting with the two particles or the atom and the field above, and the states of that LO are approximately unchanged, then we can obtain $(|e_1\rangle|e_2\rangle+|e_2\rangle|e_1\rangle)|LO\rangle$ or $(|g\rangle|0\rangle+|e\rangle|1\rangle)|LO\rangle$. Otherwise, if there is no LO interacting with the two particles, or the INT between them ceases, the entangled state disentangles into a mixed state $|e_1\rangle|e_2\rangle\langle e_1|\langle e_2|+|e_2\rangle|e_1\rangle\langle e_2|\langle e_1|$.

\section{Review of the Argument between Einstein and Bohr}

There are two focal points of the argument between Einstein \cite{A. Einstein} and Bohr \cite{Bohr}. The first is the physical reality presented by Einstein \emph{et al}.: \textquotedblleft If, without in any way disturbing a system, we can predict with certainty the value (\emph{i.e.}, with probability equal to unity) of a physical quantity, then there exists an element of physical reality corresponding to this physical quantity."  \textquoteleft Without in any way disturbing a system' means no INT with apparatus.

However, Bohr refuted the physical reality with INT between a measured particle and a measuring apparatus, and he considered the momentum exchanged between the particle and the separate parts of the apparatus.

We consider that if one knows initially some physical quantities of a system and its Hamiltonian, he or her still cannot know all the quantities such as a continuous variable momentum later due to unavoidable uncertain tiny INTs. If a system state is not disturbed or is unchanged, the measuring apparatus state should be unchanged too, which also corresponds to no measurement or no measured system existing, and then no result or information about the system can be read. So we think that Bohr won the argument on this point based on INT.

The second is the physical locality presented also by Einstein \emph{et al}.\cite{A. Einstein}. They considered that two systems become an entangled state, which is expressed by their Eq. (7), from an initial product state due to interact. They thought: \textquotedblleft At the time of measurement the two systems no longer interact, no real change can take place in the second system in consequence of anything that may be done to the first system." They emphasized that INT decides the coherence of the two systems, \emph{i.e}., entanglement.

But Bohr refuted the physical locality with his famous complementarity \cite{J. A. Wheeler}, which is unrelated to any INT. When his complementarity was used for a particle, it seemed to explain why a particle has wave-particle duality, and the retention of the entanglement after INT ceased (non-local property) for a CS.

However, we have explained that the wave property of a particle is based on INT, not by any other principle, and entangled states are also kept by INT that exchanges energy, momentum and angular momentum. As far as we know, whether an entangled state is local or non-local has not been completely proved experimentally, since the detection loophole \cite{E. Togan} and the lightcone loophole \cite{M. A. Rowe}, \emph{i.e}., the low detection efficiency of photons and the lack of a spacelike separation associated with measurement, have not been closed at the same time till 2014. The experiment of the ghost diffraction \cite{D. V. Strekalov} from entangled photons may be considered a proof of the existence of quantum non-locality, but the ghost diffraction can also be observed from un-entangled photons \cite{J. Hua}. B.-G. Englert \cite{Englert} thought that quantum mechanics is local and refuted \cite{B-G Englert} those free loophole tests of the nonlocal experiments in 2015-2017 \cite{B. Hensen, Marissa Giustina, Shalm, W. Rosenfeld} from a Bayesian View. So we think that the physical locality related to INT is right, and that Einstein therefore won the argument on the second point.

As for the famous words of Einstein that God does not play dice, we think that the probability issue is decided in principle by unavoidable uncertain tiny INTs, the Schr\"{o}dinger equation, and boundary and initial conditions. Unknown hidden variables should not be considered before one will discover all unavoidable uncertain tiny INTs in future.

\section{Conclusions}

In classical mechanics, the external force or the potential energy in a Lagrangian or Hamiltonian may cause one to forget that it comes from interaction between a considered macro-particle and another object. In non-relativistic quantum mechanics, we often concentrate our attention on a considered particle and its external potential energy, and then often forget other objects and the external potential energy being the approximation of the interaction energy belonging to a larger composite system.

If we consider those once neglected interactions between micro-systems and some macro-objects or environments, especially the latter states, and consider Newton's idea on measurement, we should modify the theory of non-relativistic quantum mechanics, can understand some quantum phenomena, and can mediate logically some contradictions to internal consistency. These can partly satisfy our logic and intuition, and then may partly enable us to perfect the theory.

We also review the arguments between Einstein and Bohr on physical reality and physical locality based on interaction, and think that the locality is right and the reality is wrong.

In order to solve issues of non-relativistic quantum mechanics, two important things may be to find unknown interactions in practical composite systems and even to find the fifth fundamental interaction, rather than finding hidden variables.

\begin{acknowledgments}

I thank Choo Hiap Oh, Berthold-Georg Englert, Gerard't Hooft, Christopher Monroe, Chin-wen Chou, Tong-Cang Li, Cheng-Zu Li, Lin-Mei Liang, Shou-Yong Pei, Su-Pen Kou, Xiao-Ming Liu, Hai-Bo Wang, Guo-Jun Jin, Dian-Min Tong, Gui-Lu Long, Shi-Dong Liang, Zhi-Bin Li, Si-Xia Yu, Guo-Yong Xiang, Yu Shi, Xue-Xi Yi, Chun Liu, Feng-Li Yan, Hao-Sheng Zeng, Ying-Hua Ji, Xin-Qi Li, Cheng-Jie Zhang, Gui-Qin Li, Li-Fan Ying, Pei-Zhu Ding, Xian-guo Jiang, Chun-Yu Chang, Li-Yu tian and my colleagues Jian Zou, Bin Shao, Xiu-San Xing, Xiang-Dong Zhang, Yu-Gui Yao, Feng Wang, Jun-Gang Li, Hao Wei, Bing-Cong Gou, Yan-Xia Xing, Fan Yang, Zhao-Tan Jiang, Wen-Yong Su, Jin-Song Miao, Chang-Hong Lu, Rui Wan, Hai-yun Hu, Jin-Fang Cai, Yan-Quan Feng, Liang Wan, Lin Li, Shao-Bo Zheng, Cheng-Cheng Liu, Hong-Kang Zhao, Yong-You Zhang, Li-Jie Shi, Xin-Bin Song, Jian-feng He, Yu-Long Liu, Yong-Jun L\"{u}, Da-Zhi Xu, Li-Da Zhang, Sheng-Li Zhang, Ye Cao, Jiang-Wei Shang, An-Ning Zhang, Fei Wang, Han-Chun Wu for discussions and comments, Jin-Song Miao for translations of early German literature, Hao Wei, Gui-Qin Li Li-Fan Ying, Hui Yan and Yong-Jun L\"{u} for helps, and Pei-Zhu Ding, Shou-Fu Pan and Xian-guo Jiang for encouragements. The work was supported by the fundamental research fund of Beijing Institute of Technology(2010-2011) and by the National Natural Science Foundation of China (Grant Nos. 11075013, 11375025, 11474020 of Bing-Cong Gou, 11674024 and 11875086).

\end{acknowledgments}

\end{document}